
File:


\font\titolino=cmbx10
\font\tsnorm=cmr10
\font\tscors=cmti10

\font\tscorsp=cmti9
\magnification=1200

\hsize=148truemm
\hoffset=10truemm
\parskip 3truemm plus 1truemm minus 1truemm
\parindent 8truemm
\newcount\notenumber

\def\note{\advance\notenumber by 1 \footnote{$^{\the\notenumber}$}}
\def\ref#1{\medskip\everypar={\hangindent 2\parindent}#1}
\def\beginref{\begingroup
\bigskip
\leftline{\titolino References.}
\nobreak\noindent}
\def\endref{\par\endgroup}
\def\beginsection #1. #2.
{\bigskip
\leftline{\titolino #1. #2.}
\nobreak\noindent}
\def\beginappendix #1.
{\bigskip
\leftline{\titolino Appendix #1.}
\nobreak\noindent}
\def\beginack
{\bigskip
\leftline{\titolino Acknowledgments}
\nobreak\noindent}

\nopagenumbers
\rightline{}
\rightline{\tscors April 1993}
\vskip 20truemm
\centerline{\titolino FUNCTIONAL CHANGE OF VARIABLES}
\bigskip
\centerline{\titolino IN THE WHEELER--DE WITT EQUATION}
\vskip 15truemm
\centerline{\tsnorm Marco Cavagli\`a}
\bigskip
\centerline{\tscorsp ISAS, International School for Advanced
Studies, Trieste, Italy}
\vfill
\centerline{\tsnorm ABSTRACT}
\begingroup\tsnorm\noindent
I present a new way to solve the Wheeler--de Witt equation using the
invariance of the classical lagrangian under reparametrization. This
property allows one to introduce an arbitrary function for each degree of
freedom of the wave function $\Psi$: this arbitrariness can be used to
fix the asymptotic behaviour of $\Psi$ so as to obtain a wave function
representing a closed universe or a wormhole. These considerations are
applied in detail to the Kantowsky--Sachs spacetime.
\vfill
\hrule
\noindent
Mail Address:
\hfill\break
ISAS-International School for Advanced Studies
\hfill\break
Via Beirut 2-4, I-34013 Miramare (Trieste)
\hfill\break
Electronic mail: 38028::CAVAGLIA or CAVAGLIA@TSMI19.SISSA.IT
\endgroup
\eject
\footline{\hfill\folio\hfill}
\pageno=1
\beginsection 1. Introduction.
Different problems arise in writing and solving the Wheeler--de Witt
(WDW) equation. An important one is how to associate the quantum
operators to the classical quantities. Generally we start with a
classical lagrangian which is a function of some number of variables: for
example in the minisuperspace formalism the euclidean action for
gravity minimally coupled to a scalar field is [1,2]
$$S_E=-{1\over 2}\int dt\bigl[a\dot
a^2+a-a^3\dot\varphi^2\bigr]\eqno(1.1)$$
and the problem is formally reduced to two degrees of freedom. To
quantize this system we must associate a representation of the quantum
operators representing the conjugate momenta to the variables $a$ and
$\varphi$, i.e. $\Pi_a$ and $\Pi_\varphi$. For instance we may put:
$$\eqalignno{&\Pi_a^2\rightarrow {1\over a^p}{\partial\over\partial a}
\biggl[a^p{\partial\over\partial a}\biggr]&(1.2a)\cr
&\Pi_\varphi^2\rightarrow {1\over\varphi^q}{\partial\over\partial\varphi}
\biggl[\varphi^q{\partial\over\partial\varphi}\biggr]&(1.2b)\cr}$$
where $p$ and $q$ are two arbitrary positive integers. Choosing $p=1$ and
$q=0$ the WDW equation assumes the form:
$$\Biggl[{1\over a}{\partial\over\partial a}
a{\partial\over\partial a}-{1\over
a^2}{\partial^2\over\partial\varphi^2}-a^2\Biggr]
\Psi(a,\varphi)=0.\eqno(1.3)$$
However, using the procedure described above we have two ambiguities: the
factor ordering of the quantum operators (i.e. $p$ and $q$) and the
choice of the variables to quantize: for instance to obtain wave
functions describing wormholes [2] a better choice is to quantize the
lagrangian written in terms of the variables
\medskip
\line{\hfil$x=a\sinh\varphi$,\hfil$y=a\cosh\varphi$.\hfil}
\medskip
\noindent
With these variables the WDW equation becomes ($p=0$, $q=0$):
$$\Biggl[{\partial^2\over\partial y^2}-{\partial^2\over\partial
x^2}-y^2+x^2\Biggr]\Psi(x,y)=0.\eqno(1.4)$$
In this way we obtain the solutions:
$$\Psi(x,y)=\psi_n(x)\psi_n(y)\eqno(1.5)$$
where $\psi_n$ represents the wave function of the one dimensional
harmonic oscillator of order $n$. In the case of more degrees of freedom
a simple choice for the variables to quantize is generally not possible:
for example in the case of a Kantowsky--Sachs spacetime (see section 4)
it is not trivial find the appropriate variable transformations in order
to obtain a definite type solution, for instance a wormhole wave
function.

In the following I will show that it is possible to quantize the system
and write the WDW equation without fixing the transformations. The form
of these can be chosen later by specifying the desired
asymptotic behaviour of the wave function. We want to emphasize that the
difference between the two methods is that in our approach we fix the
variable transformations only after quantization of the system. In this
way the quantization is independent of the transformations which can be
interpreted like ``gauge'' functions because a particular choice of them
does not change the physics.

The structure of the paper is the following: in the next section we shall
recall briefly the main aspects of the WDW equation and we will define
the notations; in the third section we shall introduce the
transformations for the general case; finally in the last section we
apply these considerations to the simple case of a Kantowsky--Sachs
spacetime.
\beginsection 2. WDW equation.
In the following let us introduce the standard notations for the WDW
equation. In particular we will write the line element in the form [3]:
$$ds^2=(N^2-N_iN^i)dt^2+2N_idx^idt+h_{ij}dx^idx^j\eqno(2.1)$$
where $N$ represents the lapse function, $N^i$ is the shift vector and
$h_{ij}$ is the three-space metric. The euclidean hamiltonian for general
relativity ($M_p$ is the Planck mass) is then [4]
$$H_E=\int(N{\cal H}_G+N_i{\cal H}^i)d^3x\eqno(2.2)$$
where (in the following ${}^{(3)}R$ is the scalar curvature tensor
relative to the space metric)
$${\cal H}_G=16\pi M_p^{-2}{\cal H}_{ijkl}\Pi^{ij}\Pi^{kl}+\sqrt
h\Biggl[\biggl({M_p^2\over16\pi}\biggr){}^{(3)}R+T_{44}\Biggr],\eqno(2.3)$$
and (``$~|~$'' represents the covariant derivative with respect to the
space metric)
$${\cal H}^i=-2\Pi^{ij}{}_{|j}-\sqrt hT^{4i}.\eqno(2.4)$$
(2.3) and (2.4) are written in terms of the superspace metric
$${\cal H}_{ijkl}={1\over 2\sqrt
h}(h_{ik}h_{jl}+h_{il}h_{jk}-h_{ij}h_{kl}),\eqno(2.5)$$
in terms of the conjugate momentum to $h_{ij}$ (${\bf K}^{ij}$ is the
second fundamental form)
$$\Pi^{ij}={M_p^2\over 16\pi}\sqrt h({\bf K}^{ij}-h^{ij}{\bf
K}),\eqno(2.6)$$
and in terms of the stress energy tensor of the matter field
$T_{\mu\nu}$. We suppose also that the cosmological constant is zero. The
lapse function and the shift vector can be considered as Lagrange
multipliers; then the classical equations of motion can be written:
$$\eqalignno{&{\cal H}_G=0,&(2.7a)\cr
&{\cal H}^i=0.&(2.7b)\cr}$$
The quantization of the problem can be achieved identifying the classical
quantities $\Pi^{ij}$ with the operators [3]
$$\Pi^{ij}\rightarrow -\biggl({M_p^2\over 16\pi}\biggr)
{\delta\over\delta h_{ij}}\eqno(2.8)$$
which satisfy the euclidean commutation relations:
$$\eqalign{&\bigl[h_{ij},h_{kl}\bigr]=0,\cr
&\bigl[\Pi_{ij},\Pi_{kl}\bigr]=0,\cr
&\bigl[h_\alpha,\Pi_\beta\bigr]={M_p^2\over
16\pi}\delta_{\alpha\beta},\hbox to 10truemm{}\alpha=(ij),\hbox to
10truemm{}\beta=(kl).\cr}\eqno(2.9)$$
and analogously for the classical momenta of the matter fields. With this
identification we obtain from $(2.7a)$ the WDW equation describing the
quantum properties of the gravitational field coupled to matter:
$$\Biggl[{\cal H}_{ijkl}{\delta^2\over\delta h_{ij}
\delta h_{kl}}+\sqrt h\biggl[{}^{(3)}R+{16\pi\over M_p^2}T_{44}\biggr]\Biggr]
\Psi(h_{ij},\phi)=0\eqno(2.10)$$
where now the stress energy tensor of the matter fields $\phi$ is a
quantum operator and $\Psi$ represents the wave function of the system.
{}From $(2.7b)$ we obtain the equation:
$$\Biggl[-2\biggl({M_p^2\over 16\pi}\biggr)\biggl[{\delta\over
\delta h_{ij}}\biggr]_{|j}+\sqrt hT^{4i}\Biggr]\Psi(h_{ij},\phi)=0\eqno(2.11)$$
that represents a constraint identically satisfied by the wave function.
\beginsection 3. Transformations for the general WDW equation.
Let us come back to the line element (2.1) and perform the
transformation:
$$h_{ij}\rightarrow h'_{ij}=f^{-1}_{ij}(h_{kl}).\eqno(3.1)$$
We stress that (3.1) is not a change of coordinates in the metric but a
functional transformation of the space metric. To make this consideration
explicit we can consider, for example, the line element
$$ds^2=dt^2+a^2(t)d\vec x^2\eqno(3.2)$$
which corresponds to (2.1) with $N=1$, $N_i=0$ and
$h_{ij}=a^2(t)\delta_{ij}$. A transformation (3.1) could be
$$a^2(t)=b(t).\eqno(3.3)$$
In this case we do not change the coordinates, only the functional form
of the scale factor. Let us now substitute (3.1) into the lagrangian: we
can calculate the hamiltonian with respect to the new space metric and
conjugate momentum. This latter will be related to the previous
expression (2.6) by the following relation:
$$(\Pi^{ij})'=\Pi^{kl}{\partial f_{kl}\over\partial h'_{ij}}.\eqno(3.4)$$
Now we can repeat all the steps described in the previous paragraph to
find the WDW equation. (2.8) becomes
$$(\Pi^{ij})'\rightarrow -\biggl({M_p^2\over 16\pi}\biggr)
{\delta\over\delta h'_{ij}}\eqno(3.5)$$
which implies that the quantum operator that must be substituted for
$\Pi^{ij}$ in (2.3) is
$$\Pi^{ij}\rightarrow -{\partial f^{-1}_{kl}\over\partial h_{ij}}
\biggl({M_p^2\over 16\pi}\biggr){\delta\over\delta h'_{kl}}.\eqno(3.6)$$
We can easily note that the WDW equation is now formally different
from the previous expression (2.10) because of the presence of the
factors
$${\partial f^{-1}_{kl}\over\partial h_{ij}}$$
which represent the arbitrariness in the choice of the variables to be
quantized. The functions $f_{ij}$ can then be interpreted as ``gauge''
functions and can be fixed to obtain an appropriate form for the WDW
equation.
\beginsection 4. Kantowsky--Sachs spacetime.
Now we will apply the considerations of the latter paragraph to a
particular ansatz. The simplest non trivial case is obtained when the
gravitational lagrangian depends on two degrees of freedom; we assume
then the spacetime manifold to be described by the Kantowsky--Sachs line
element:
$$ds^2=N^2(t)dt^2+a^2(t)d\chi^2+b^2(t)d\Omega_2^2\eqno(4.1)$$
where $\chi$ is the coordinate of the 1-sphere, $0\le\chi<2\pi$,
$d\Omega_2^2$ represents the line element of the 2-sphere and $N(t)$ is
the lapse function. In the pure gravity case this problem was first
solved by Fishbone [5] using the ``exponential route'' choice of
variables [6]. Because the WDW equation is independent of the lapse
function we can simplify the line element choosing $N=1$. The euclidean
action of our problem is
$$S_E=\int_\Omega d^4x\sqrt g\Biggl[-{M_p^2\over 16\pi}R+L(\phi)\biggr]
+\int_{\partial\Omega} d^3x\sqrt h {M_p^2\over 8\pi}{\bf K}\eqno(4.2)$$
where $\Omega$ is the compact four dimensional manifold described by
(4.1), $R$ is the curvature scalar, $L(\phi)$ is the matter lagrangian,
${\bf K}$ is the trace of the extrinsic curvature of the boundary
$\partial\Omega$ of $\Omega$ and $h$ is the determinant of the induced
metric over $\partial\Omega$.

Let us discuss first the pure gravity case. Substituting (4.1) in (4.2)
we obtain:
$$S_E=-\int dt\bigl[a\dot b^2+2\dot ab\dot b+a\bigr]\eqno(4.3)$$
where the dot represents differentiation with respect to $t$ and we have
integrated over the coordinates of the 1-sphere and 2-sphere. The problem
is then formally reduced to two degrees of freedom: $a$ and $b$. From
(4.3) it is easy to derive the classical equations for $a$ and $b$
varying with respect to the two scale factors [7]:
$$\eqalignno{2{\ddot b\over b}+{\dot b^2\over b^2}-{1\over b^2}&=0,&(4.4a)\cr
{\ddot b\over b}+{\ddot a\over a}+{\dot a\dot b\over ab}&=0.&(4.4b)\cr}$$
Now we look for the quantum solutions. To associate the classical
quantities $a$ and $b$ and their momenta to quantum operators we need a
new form for the action (4.3): this is necessary because the lagrangian
is not quadratic in the canonical momenta. Following the prescriptions of
the previous section we introduce two functions in order to
separate the canonical momenta:
$$\eqalignno{a&=f(x,y),&(4.5a)\cr
b&=g(x,y).&(4.5b)\cr}$$
Substituting into the lagrangian we can require that the term
proportional to $\dot x\dot y$ vanishes; if $f$ does not depend on $y$
this condition implies that the functions $f$ and $g$ fulfill
$$f{\partial g\over\partial x}+g{\partial f\over\partial x}=0.\eqno(4.6)$$
Analogous results can be obtained with the conditions $f(x,y)\equiv
f(y)$, or $g(x,y)\equiv g(x)$, or $g(x,y)\equiv g(y)$. From (4.6) we
obtain the final form for $a$ and $b$:
$$\eqalignno{a&=f\equiv f(x),&(4.7a)\cr
b&=g={h(y)\over f(x)},&(4.7b)\cr}$$
where $h(y)$ is an arbitrary function of $y$. Substituting (4.7) into the
lagrangian we can write the classical hamiltonian in terms of the new
canonical variables $x$ and $y$ and their canonical momenta:
$$H=-{f\over 4h^2}\Biggl[{1\over(h'/h)^2}\Pi_y^2
-{1\over(\dot f/f)^2}\Pi_x^2-4h^2\Biggr].\eqno(4.8)$$
Now we can quantize the system substituting for the classical quantities
quantum operators. The WDW equation (2.10) becomes:
$$\Biggl[{1\over(h'/h)^2}{\partial^2\over\partial y^2}-
{1\over(\dot f/f)^2}{\partial^2\over\partial
x^2}-4h^2\Biggr]\Psi(x,y)=0,\eqno(4.9)$$
where now the dot represents differentiation with respect to $x$ and the
prime denotes differentiation with respect to $y$. We can easily solve
(4.9) putting
$$\Psi(x,y)=\chi(x)\sigma(y).\eqno(4.10)$$
With this substitution (4.9) separates into the following two equations:
$$\eqalignno{{d^2\chi\over dx^2}&=K\biggl({\dot f\over
f}\biggr)^2\chi,&(4.11a)\cr
{d^2\sigma\over dy^2}&=(K+4h^2)\biggl({h'\over h}\biggr)^2
\sigma,&(4.11b)\cr}$$
where $K$ is an arbitrary constant. We want to discuss the asymptotic
behaviour of (4.10) in terms of the old variables $a$ and $b$. Using
(4.7) we can rewrite (4.11) in the following form:
$$\eqalignno{&{d^2\chi\over df^2}+{\ddot f\over\dot f^2}{d\chi\over
df}-{K\over f^2}\chi=0,&(4.12a)\cr
&{d^2\sigma\over dh^2}+{h''\over h'^2}{d\sigma\over
dh}-\biggl(4+{K\over h^2}\biggr)\sigma=0.&(4.12b)\cr}$$
Now we can fix the asymptotic behaviour of the wave functions in $a$ and
$b$ simply fixing their behaviour in $f$ and $h$. Moreover the asymptotic
behaviours can be chosen imposing suitable conditions for these two
functions. Let us see some examples.

If we want to obtain a solution damped for large values of $a$ and $b$ we
can choose $K=\nu^2>0$ and fix the ``gauge'' in the following way:
$${h''\over h'^2}={1\over h}.\eqno(4.13)$$
With this condition $(4.12b)$ becomes:
$${d^2\sigma\over dh^2}+{1\over h}{d\sigma\over
dh}-\biggl(4+{\nu^2\over h^2}\biggr)\sigma=0.\eqno(4.14)$$
Then the solutions of equation (4.14) are
$$\sigma=Z_\nu(2ih)=Z_\nu(2iab)\eqno(4.15)$$
where $Z_\nu$ is any Bessel function of index $\nu$. With $K=\nu^2$
$(4.12a)$ becomes:
$${d^2\chi\over df^2}+{\ddot f\over\dot f^2}{d\chi\over
df}-{\nu^2\over f^2}\chi=0.\eqno(4.16)$$
Choosing $\ddot f=0$ we obtain:
$${d^2\chi\over df^2}-{\nu^2\over f^2}\chi=0\eqno(4.17)$$
which has the non singular solution
$$\chi=\sqrt f\cdot f^{\sqrt{1+4\nu^2}/2}.\eqno(4.18)$$
Finally we can write the complete form of the wave function
$$\Psi(a,b)=\sqrt a\cdot a^{\sqrt{1+4\nu^2}/2}Z_\nu(2iab).\eqno(4.19)$$
If we want to obtain the transformations between $a$, $b$ and $x$, $y$
that give rise to the solution (4.19) we can solve (4.13) and the
corresponding equation for $f$. We obtain:
$$\eqalign{&ab=C_1\cdot\exp(C_2\cdot y),\cr
&a=C_3+C_4 x,}\eqno(4.20)$$
where $C_1$, $C_2$, $C_3$ and $C_4$ are arbitrary constants. (4.20)
represent the transformations (3.1) in our case. Choosing for simplicity
$C_1=C_2=C_4=1$ and $C_3=0$ we obtain that the line element (4.1) can be
written in the following way:
$$ds^2=dt^2+x^2(t)d\chi^2+{e^{2y(t)}\over x^2(t)}d\Omega_2^2.\eqno(4.21)$$
As we have mentioned, the choice of the asymptotic behaviour of the wave
function (4.19) for large values of the scale factor $b$ can be:
$$\Psi\approx\exp(-ab).\eqno(4.22)$$
This is exactly the asymptotic behaviour of a solution representing a
wormhole with spatial symmetry $S^2\times S^1$ [8]; then (4.19) can be
interpreted as describing a quantum wormhole.

In the same way we can discuss the case with $L(\phi)\not=0$. Let us see
the cases of scalar and electromagnetic fields. If we add a scalar field
the action (4.3) becomes
$$S_E=-\int dt\bigl[a\dot b^2+2\dot ab\dot b+a-ab^2\dot\Phi^2\bigr]
\eqno(4.23)$$
and the WDW equation assumes the form
$$\Biggl[{\partial^2\over\partial\Phi^2}-{1\over(h'/h)^2}
{\partial^2\over\partial y^2}+{1\over(\dot f/f)^2}
{\partial^2\over\partial x^2}+4h^2\Biggr]\Psi(x,y,\Phi)=0.\eqno(4.24)$$
The matter degree of freedom can be easily separated by putting
$\Psi=\zeta(\Phi)\xi(x,y)$. With this substitution (4.24) separates into
$$\eqalignno{&{d^2\zeta\over d\Phi^2}+\omega^2\Phi=0,&(4.25a)\cr
&\Biggl[-{1\over(h'/h)^2}
{\partial^2\over\partial y^2}+{1\over(\dot f/f)^2}
{\partial^2\over\partial x^2}+4h^2-\omega^2\Biggr]\xi(x,y)=0.&(4.25b)\cr}$$
$(4.25b)$ can be separated to obtain
$$\eqalignno{{d^2\chi\over dx^2}&=(\omega^2+K)\biggl({\dot f\over
f}\biggr)^2\chi,&(4.26a)\cr
{d^2\sigma\over dy^2}&=(K+4h^2)\biggl({h'\over h}\biggr)^2
\sigma.&(4.26b)\cr}$$
We can easily solve $(4.25a)$ and (4.26) to get the wave function. In the
case $K=\nu^2$ we find:
$$\Psi(a,b,\Phi)=\sqrt a\cdot
a^{\sqrt{1+4(\nu^2+\omega^2)}/2}Z_\nu(2iab)e^{\pm i\omega\Phi}.\eqno(4.27)$$
Now let us examine the case with an electromagnetic field. If we choose
for the vector potential the ansatz [7]
$$A_\mu=A(t)\delta_{\chi\mu},\eqno(4.28)$$
the action (4.3) becomes:
$$S_E=-\int dt\biggl[a\dot b^2+2\dot ab\dot b+a-{b^2\over a}\dot A^2\biggr]
\eqno(4.29)$$
and the corresponding WDW equation is:
$$\Biggl[{\partial^2\over\partial A^2}+{1\over f^2}\biggl[-{1\over(h'/h)^2}
{\partial^2\over\partial y^2}+{1\over(\dot f/f)^2}
{\partial^2\over\partial x^2}+4h^2\biggr]\Biggr]
\Psi(x,y,A)=0\eqno(4.30)$$
which is easily separated using an ansatz similar to the previous case.
In the case $K=\nu^2$ we obtain the solutions:
$$\eqalignno{\Psi(a,b,A)&=\sqrt a
Z_\alpha(i\omega a)Z_\nu(2iab)e^{\pm i\omega A},&(4.31a)\cr
\Psi(a,b,A)&=Z_\nu(i\omega a)Z_\nu(2iab)e^{\pm i\omega A},&(4.31b)\cr}$$
where $\alpha=\pm\sqrt{1+4\nu^2}/2$ and we have used (4.13) for
$h$ and respectively $\ddot f=0$ and $\ddot f/\dot f^2=1/f$ for
$f$.
\beginack
I am very grateful to Prof. V. de Alfaro for interesting discussions. I
thank also Prof. J. Nelson and Prof. P. Haines.
\vfill\eject
\beginref
\ref [1] Hawking, S.W. and D.N. Page 1990, {\tscors Phys. Rev.} D {\bf
42}, 2655.
\ref [2] Hawking, S.W. 1990, {\tscors Mod. Phys. Lett.} A {\bf 5}, 453.
\ref [3] Hawking, S.W. 1984, {\tscors Nucl. Phys.}  B {\bf 239}, 257.
\ref [4] Hawking, S.W. and D.N. Page 1986, {\tscors Nucl. Phys.} B {\bf
264}, 185.
\ref [5] MacCallum, M.A.H. in: {\tscors Quantum Gravity: an Oxford
Symposium}, eds. by C.J. Isham, R. Penrose and D.W. Sciama, 1975.
\ref [6] Brill, D. and R.H. Gowdy 1970, {\tscors Repts. Prog. Phys.} {\bf
33}, 413.
\ref [7] Cavagli\`a M., V. de Alfaro and F. de Felice, 1993, {\tscors
Torino University Preprint} {\bf DFTT 2/93}.
\ref [8] Campbell L.M. and L.J. Garay, 1991, {\tscors
Phys. Lett.} B {\bf 254}, 49.
\endref
\vfill\eject
\bye